\documentclass{sigchi-ext}
\usepackage[T1]{fontenc}
\usepackage{textcomp}
\usepackage[scaled=.92]{helvet} % for proper fonts
\usepackage{graphicx} % for EPS use the graphics package instead
\usepackage{balance}  % for useful for balancing the last columns
\usepackage{booktabs} % for pretty table rules
\usepackage{ccicons}  % for Creative Commons citation icons
\usepackage{ragged2e} % for tighter hyphenation

\def\plaintitle{More Specificity, More Attention to Social Context: Reframing How We Address ``Bad Actors''} 
\def\emptyauthor{}
\def\plainkeywords{feminism, harassment, online communities}

\title{More Specificity, More Attention to Social Context: Reframing How We Address ``Bad Actors''}

\numberofauthors{1}
\author{%
  \alignauthor{%
    \textbf{Libby Hemphill}\\
    \affaddr{University of Michigan} \\
    \affaddr{Ann Arbor, MI 48104, USA} \\
    \email{libbyh@umich.edu} }\alignauthor{%
    } }

\definecolor{linkColor}{RGB}{6,125,233}
\hypersetup{%
  pdftitle={\plaintitle},
  pdfauthor={\emptyauthor},
  pdfkeywords={\plainkeywords},
  bookmarksnumbered,
  pdfstartview={FitH},
  colorlinks,
  citecolor=black,
  filecolor=black,
  linkcolor=black,
  urlcolor=linkColor,
  breaklinks=true,
}

\begin{document}

%% For the camera ready, use the commands provided by the ACM in the Permission Release Form.
%\CopyrightYear{2007}
%\setcopyright{rightsretained}
%\conferenceinfo{WOODSTOCK}{'97 El Paso, Texas USA}
%\isbn{0-12345-67-8/90/01}
%\doi{http://dx.doi.org/10.1145/2858036.2858119}
%% Then override the default copyright message with the \acmcopyright command.
%\copyrightinfo{\acmcopyright}

\maketitle

% Uncomment to disable hyphenation (not recommended)
% https://twitter.com/anjirokhan/status/546046683331973120
\RaggedRight{} 

% Do not change the page size or page settings.
\begin{abstract}
To address ``bad actors'' online, I argue for more specific definitions of acceptable and unacceptable behaviors and explicit attention to the social structures in which behaviors occur.
\end{abstract}



\section{Introduction}

Harassment and related undesirable behaviors online are incredibly problematic. We also know that harassment harms victims, affects an unacceptable number of internet users, disproportionately impacts members of marginalized and at-risk populations, and comes in many forms. However, efforts to curb harassment --- e.g., crowdsourced moderation, block lists, reporting and escalation to platform moderation and policy teams --- are often ineffective for victims and labor-intensive and traumatic for moderators. 

To move toward more effective methods to curb harassment, I propose that we think about the problem of harassment differently in two ways. First, we must be more specific and explicit about the behaviors and content that are unacceptable in particular contexts so that we can design targeted mechanisms for addressing them and recognize the potential unintended impacts of our interventions. Second, we must treat harassment as a social problem, not just an individual one, which demands that we address the contexts in which harassment occurs. These two shifts in the way we think about harassment come from a feminist orientation that requires us to consider power and oppression when trying to understand why and how something --- like harassment --- happens. 

I argue that harassment is, at it's root, about power --- especially the power to make someone else feel powerless. This means that I see harassment not as an issue of compliance but as a result of structural differences in power that make it possible for some actors --- maybe we call them ``bad actors'' --- to harm others. Recognizing that structural inequalities enable harassment does not mean harassers are not responsible for their actions, but it does mean that addressing ``bad actors'' requires systemic change and not just interventions targeted at individuals.

I'm interested in how social media can be leveraged to challenge existing power structures, and part of my research agenda aims to automatically identify and address unacceptable behavior. I'm a queer, white, cis woman who talks to other people on the internet and who studies online conversations. I want to reduce unacceptable behaviors because they disproportionately silence voices I think we should hear (e.g., women, racial and ethnic minorities, LGBTQ+). I mention all of this to address the requests for additional information in the workshop papers call. Next I expand on the reframings I called for and briefly propose changes they can enable.

\section{Defining Specific Unacceptable Behaviors}

Holding people to community standards requires that we have some in the first place, and it helps if they are clearly articulated and consistently applied. Unfortunately, humans don't readily or consistently agree about what behaviors are acceptable, or which behaviors are acceptable in which contexts. Instead, we have fuzzy, flexible, leaky bins into which we place and move things we witness or do ourselves; the policies we write exhibit similar properties \cite{Pater2016-bv}.

Even when we use automated or algorithmic approaches to do the classifying, classification is a fundamentally human activity, and that means that it occurs in the social, historical, technical, racial, gendered, moral etc. contexts of all other human activities. Bowker and Star \cite{Bowker1997-dh} also point out that classifying involves negotiation and that no classification scheme works for everyone. By calling for specificity, I am not suggesting that we will produce perfect, unproblematic classification of ``good'' and ``bad'' behaviors. Instead, I'm suggesting that the process of trying to articulate specific definitions will better equip us to attend to them productively. It's important for mechanism designers to articulate the problems they are trying to solve even if a fuzzy definition produces ``good enough'' results because the clarity helps us understand \textit{why} a particular mechanism works. Knowing why puts us in better positions to address issues such as workarounds bad actors develop and shifts in the communities' norms. 

\subsection{Acceptable Does Not Mean Civil}

What counts as acceptable behavior depends on who's behaving, who's seeing the behavior, where the discussion is happening, and what's being discussed. One reason these things matter is that the realities of oppression mean that rights to speech and harms from speech are inequitably distributed. Flattening behaviors into categories like acceptable and unacceptable ignores the importance of behaviors that are productively and purposefully disruptive. For example, one recent turn in research on fighting harassment uses the language of ``civility'' to describe desired behaviors. When we call for civility we are often effectively silencing marginalized voices through tone policing\footnote{http://www.robot-hugs.com/tone-policing/}. We mask our discomfort with the emotional impact of oppression in our calls for a particular type of discourse. When we set rules for discourse, we exercise power to determine what is ``real'' or `'`right'' \cite{Butler1993-hx,Foucault1981-pq}, and blunt instruments like blocking profanity to encourage civility are abuses of that power that disproportionately impact those who experience multiple oppressions that elicit emotional responses. Emotional responses are valid and informative and should be valued.  

Profanity provides an example for how clarity could help us design better systems. The presence of profanity is highly predictive of undesirable content~\cite{Martens2015-bl}. However, not all ``fucks'' are the same. Some are threats of violence\footnote{https://femfreq.tumblr.com/post/109319269825/one-week-of-harassment-on-twitter}, and some are part of valid emotional responses to external threats\footnote{https://twitter.com/rosemcgowan/status/917844865806778368}. Most classifiers can't tell the difference, and most moderators can't either if they don't have the full context of the expletive. A system that could, however, tell the difference would be more inclusive and empowering that the status quo.

\section{Recognizing Harassment as a Social Problem}

Speaking of context, the second way I argued we need to rethink harassment is to consider it a social problem instead of an individual one. Researchers often mention properties and/or features of platforms when discussing the prevalence of bad actions. For instance, reddit's karma system (like other vote-based systems) is easily gamed, encourages reposts that migrate content across communities, and reifies the dominant culture's values through social processes such as herding \cite{Massanari2015-jq,Muchnik2013-xq}. Twitter's character limits effectively discourage nuance and extended explanation, making it hard to contextualize comments or to provide enough information for outsiders to make sense of individual comments. Removing limits makes more space for hate but also for context and background knowledge that can make conversations more productive \cite{Thelandersson2014-fx}. Anonymity is a similarly multi-edged sword that is sometimes used for evil \cite{Dinakar2011-jr} and sometimes for good \cite{Andalibi2016-bz} or otherwise productively \cite{Cross2014-ry}.

Recent conversations around \#MeToo \cite{Gilbert2017-uq} and collective efforts like Time's Up\footnote{https://www.timesupnow.com/} are bringing the notion of ``harassment as a systemic issue'' into the mainstream, and it's time we do so in system design as well. Just as firing Harvey Weinstein didn't suddenly make the Weinstein Company a great place to work or stop sexual exploitation in Hollywood, playing whack-a-mole with individuals online will not end harassment. Instead, we need to recognize that platforms' affordances facilitate bad actions, that bad actions migrate across platforms or occur simultaneously in multiple spaces, that bad actions are contagious \cite{Cheng2017-qp,Phillips2017-ep}, and that features can silence, protect, and empower different groups at the same time. The systematic devaluing of marginalized voices and their experiences happens online as well as off, and we cannot stop harassment without attending to the structures and values that enable it. 

For instance, instead of designing for civility or compliance, we could design to fight oppression \cite{Smyth2014-it} or to encourage productive discussion \cite{Thelandersson2014-fx, Massanari2015-jq}. We could highlight the controversial instead of the merely popular. We could focus on encouraging learning, constructive critique, accessibility, and inclusion like fan fiction sites \cite{Fiesler2016-cp,Campbell2016-di} or well-moderated discussions such as \textbackslash r \textbackslash AskHistorians\footnote{http://reddit.com/r/AskHistorians} or Autostraddle\footnote{http://www.autostraddle.com}.

\section{Conclusion}

In order for ``bad actors'' to ``comply with community standards''\footnote{All quotes in this paragraph are from the workshop call at http://understandingbadactors.org/.}, we need to articulate them. To ``help them discover more appropriate ways of connecting with others online'', we must communicate and reify the value of connection. And to ``design more effective interventions'' we must be specific about what we're trying to accomplish and address the contexts in which both behaviors and interventions operate. I don't have clear solutions for the problems I've raised, and I don't think the way forward is technically or socially easy by any means. Instead, it's likely messy, fraught, and computationally pretty hard but will be worth it.

\balance{} 

\bibliographystyle{SIGCHI-Reference-Format}
\bibliography{bad_actors}

\end{document}